\documentclass[prl,aps,twocolumn,superscriptaddress]{revtex4-2}

\usepackage{graphicx}
\usepackage{amsmath}
\usepackage{amssymb}
\usepackage{ifthen}
\usepackage{booktabs}

\usepackage{graphicx}
\usepackage{dcolumn}
\usepackage{bm}
\usepackage{longtable}
\usepackage{gensymb}

\newcommand{\bb}{\begin{equation}}
\newcommand{\ee}{\end{equation}}
\newcommand{\ba}{\begin{eqnarray*}}
\newcommand{\ea}{\end{eqnarray*}}
\newcommand{\rhor}{\rho({\bf r})}

\newcommand{\rr}{{\mathbf r}}

\begin{document}

\title{Beyond capillary condensation: shear-induced bridging transitions in patterned slits}

\author{Alexandr \surname{Malijevsk\'y}}
\email{malijevsky@icpf.cas.cz}
\affiliation{Research group of Molecular and Mesoscopic Modelling, The Czech Academy of Sciences, Institute of Chemical Process Fundamentals, 165 02
	Prague, Czech
	Republic}
\affiliation{Department of Physical Chemistry, University of Chemistry and Technology Prague, 166 28 Prague, Czech Republic}
\author{Andrew O. \surname{Parry}}
\affiliation{Department of Mathematics, Imperial College London, London SW7 2BZ, UK}

\author{Ji\v r\'\i \hspace{0.001cm} \surname{Janek}}
\affiliation{Research group of Molecular and Mesoscopic Modelling, The Czech Academy of Sciences, Institute of Chemical Process Fundamentals, 165 02
	Prague, Czech Republic}

\begin{abstract}
\noindent 
We study the equilibrium phase behaviour of a fluid confined in a slit made from two patterned walls. Shearing the walls frustrates the fluid, due to a competition 
between capillary condensation and interface delocalisation, forcing the formation of bridging phases with different pinning properties. This leads to an unusually rich phase diagram, displaying first-order and continuous phase transitions, depending sensitively on the slit width and shear. Generalised Kelvin equations determine the phase boundaries, while the bridging phases are characterised by large correlation lengths, predictions for which are tested using a microscopic density functional model.
\end{abstract}

 \maketitle
 
Confinement can significantly modify fluid phase  behaviour
\cite{Rowlinson2002, Henderson2009, Fisher1981, Nakanishi1983, Evans1990, Kerle1996, Gelb2000, Bonn2009, Yang2020}. A familiar example is capillary condensation whereby a low-density gas, confined in a slit of width $L$, condenses to liquid at a pressure $p_{cc}$ below the pressure $p_{\rm sat}$ of bulk saturation. The pressure shift, $\delta p_{cc}=p_{\rm sat}-p_{cc}$, is then accurately described by the macroscopic Kelvin equation \cite{Thomson1871}
\begin{equation}
	\delta p_{cc}=\frac{2\gamma\cos\theta}{L}\,,
	\label{kelvin}
\end{equation}
where $\gamma$ is the liquid-gas surface tension and $\theta$ is the contact angle at the walls, identified from Young’s equation. 

Another effect arises when the confining walls exhibit different contact angles \cite{degennes83, Parry1990, Swift1991, Binder2003, Binder2008, Malijevsky2016}. If these have competing wetting properties, one with contact angle $\theta$ and the other $\pi-\theta$, the pressure shift vanishes, $\delta p_{cc}=0$, but the role of wetting is much more prominent. Capillary condensation is then better understood as interface localisation-delocalisation, where phase coexistence only occurs below the wetting/drying temperature and is suppressed when one wall is wet ($\theta=0$) and the other dry ($\theta=\pi$). In this regime the interface is weakly localised near the slit centre, with fluctuations characterised by a large parallel correlation length, which grows with the slit width $L$, which has been observed experimentally \cite{Kerle1996}.

 \begin{figure}
	\includegraphics[width=0.5\textwidth]{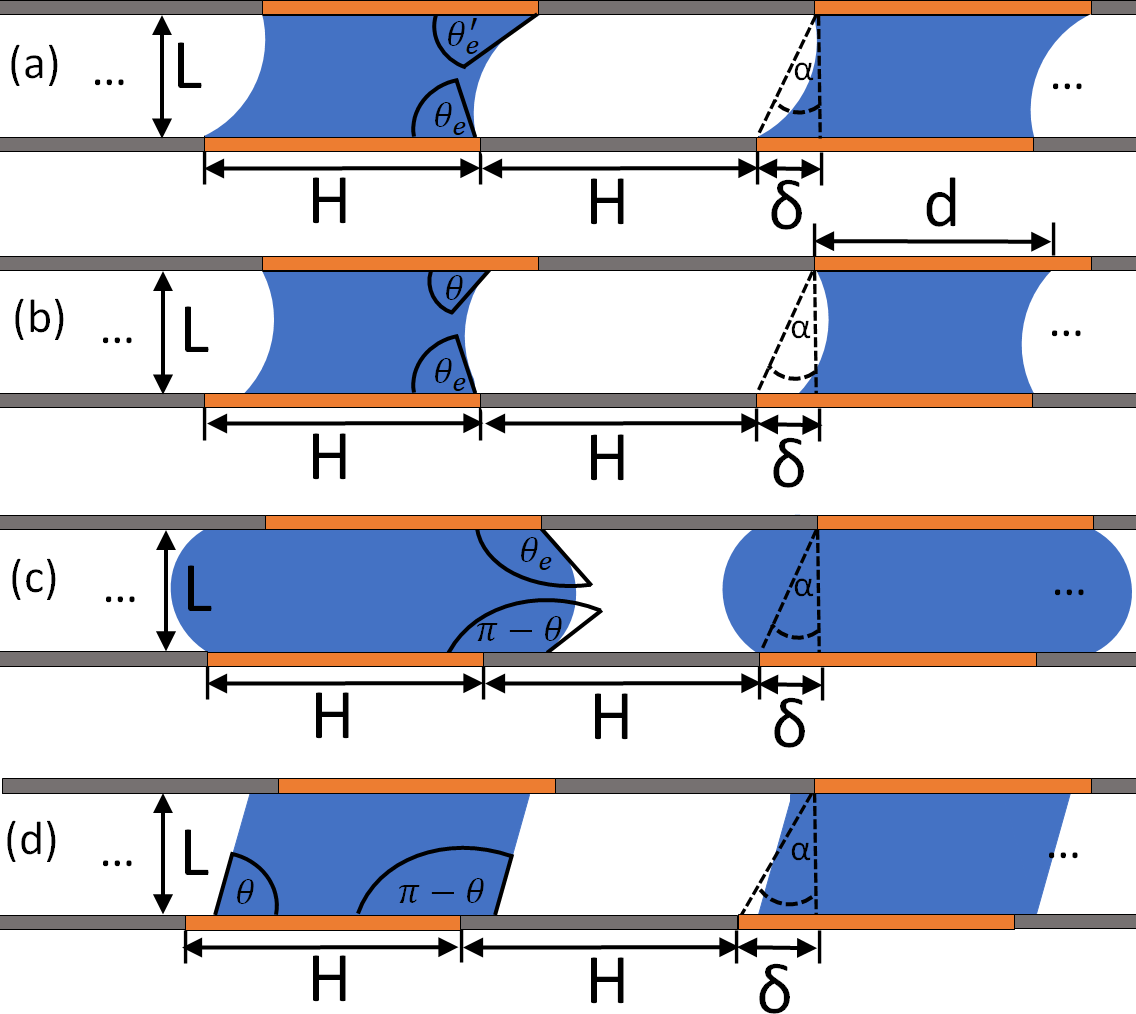}
	\caption{A patterned slit of width $L$ with alternating stripes of competing wettability, A (orange) and B (gray), each of width $H$, laterally displaced by $\delta$. (a) Type I bridging: both menisci are pinned at the patch edges and meet the walls at edge contact angles $\theta_e$ and $\theta_e'=\theta_e-2\alpha$. (b) Type IIA bridging for $p<p_{\rm sat}$: one end of the meniscus is pinned and the other is unpinned meeting patch A with contact angle $\theta$, so that the bridge covers only a length $d\le H$. (c) Type-IIB bridging,  for $p>p_{\rm sat}$, where  the unpinned end meets patch B at angle $\pi-\theta$. (d) Type III, for $p=p_{\rm sat}$, where the menisci are planar whose ends are unpinned and meet patch A at contact angle $\theta$ and patch B at contact angle $\pi-\theta$. }
\end{figure}

In this Letter we describe a capillary slit where the phase behaviour is suprisingly rich due to the competition between capillary condensation and interface delocalisation. Consider that the capillary walls are a parallel array of stripes of materials A and B -- such chemically heterogeneous confinements can now be fabricated even at the nanoscale \cite{Bruschi2017}. We suppose that  the stripes have the same width $H$, with competing contact angles $\theta_A=\theta $ and $\theta_B=\pi-\theta$. This is readily achieved in Ising model studies by reversing the sign of the surface field in regions A and B. The stripes on adjacent walls are parallel, maintaining one direction of translational invariance (say, $y$), but are sheared by a displacement $\delta$ which lies in the interval $-H<\delta<H$ (see Fig.~1). Thus, $\delta=0$ means the opposing stripes are in phase, which favours local condensation, while $\delta=\pm H$ means they are out of phase with competing wetting properties, favouring local interface delocalisation.

\begin{figure*}
	\includegraphics[width=0.32\textwidth]{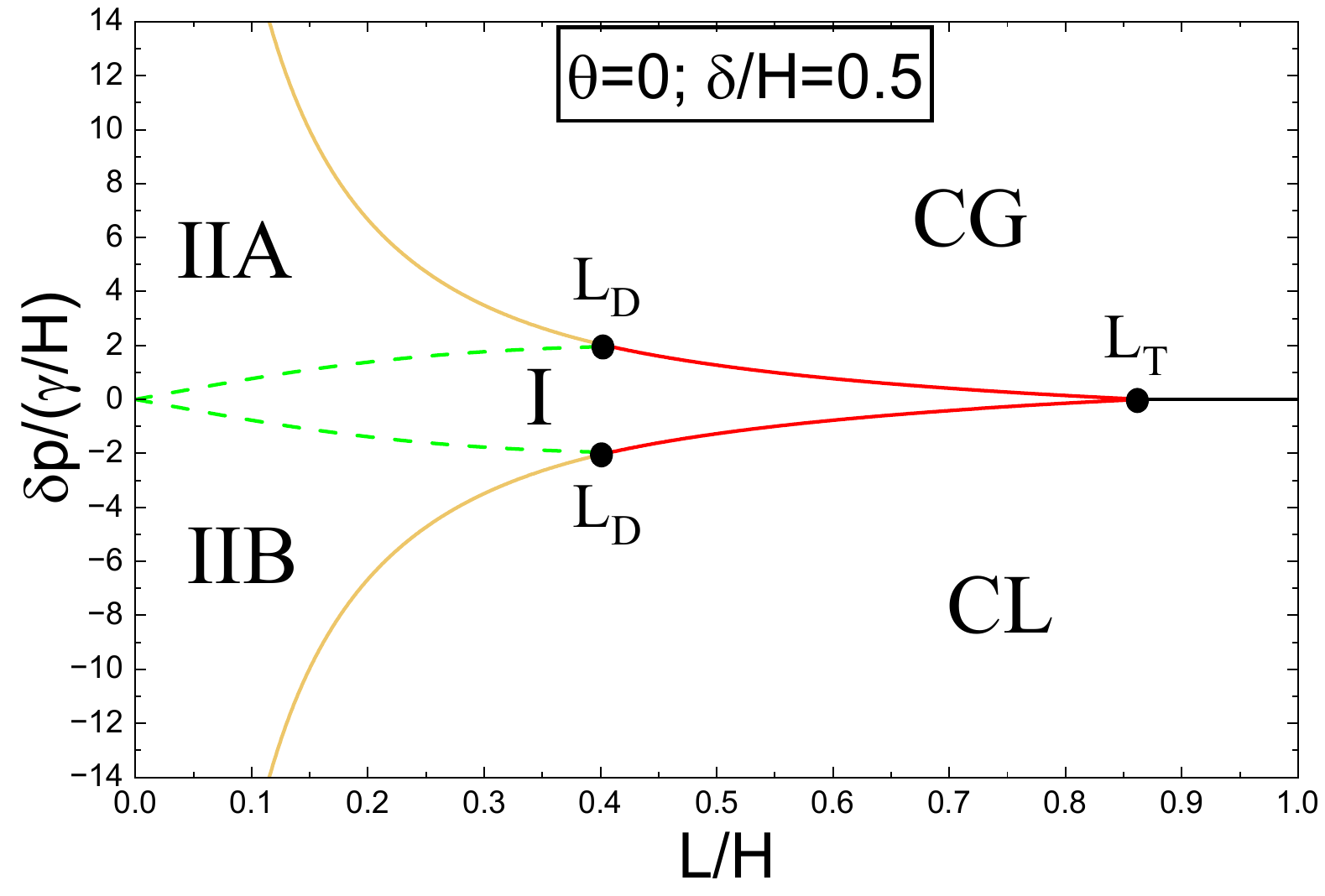}\hspace*{0.1cm} \includegraphics[width=0.32\textwidth]{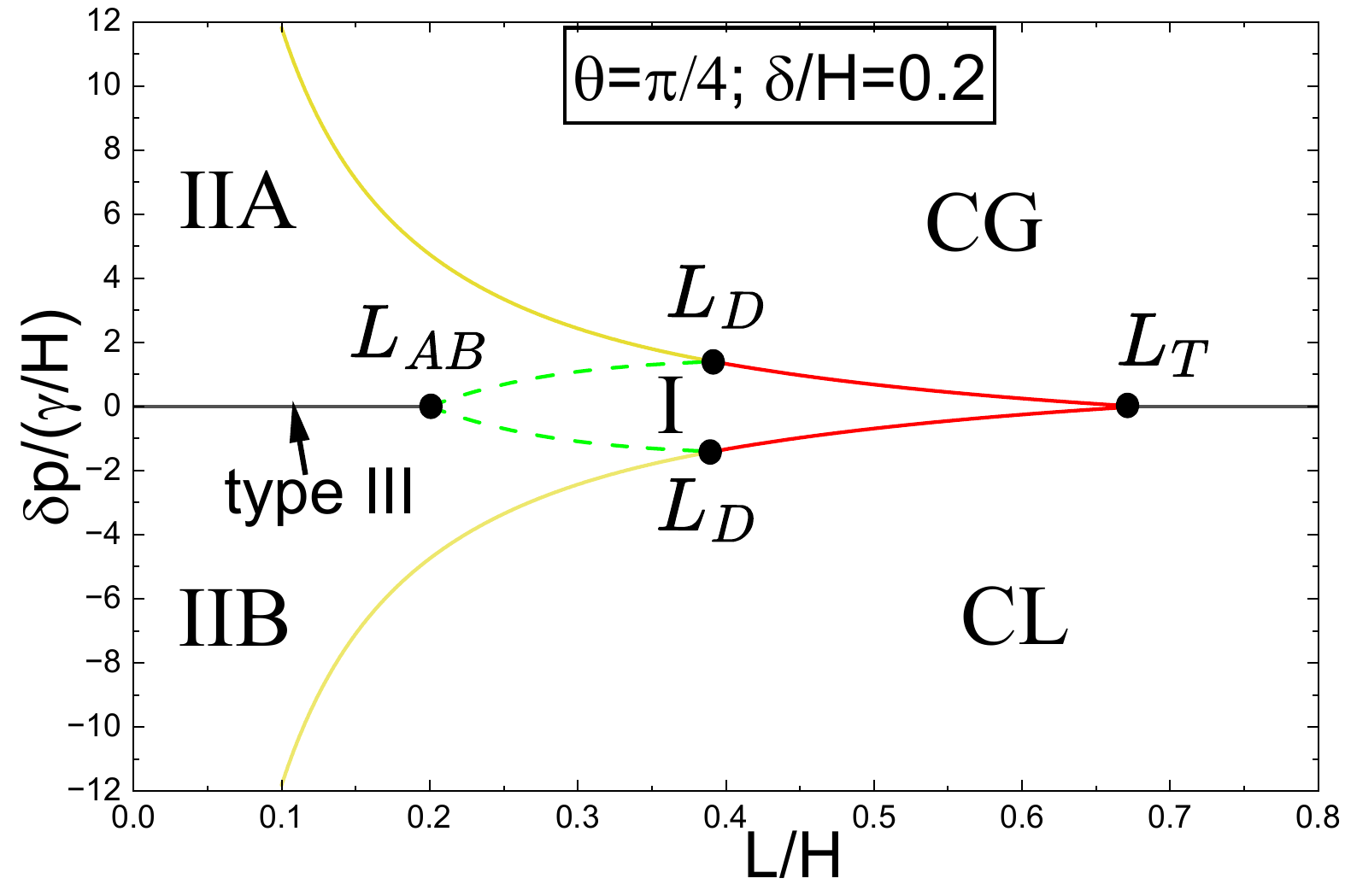}
	\includegraphics[width=0.32\textwidth]{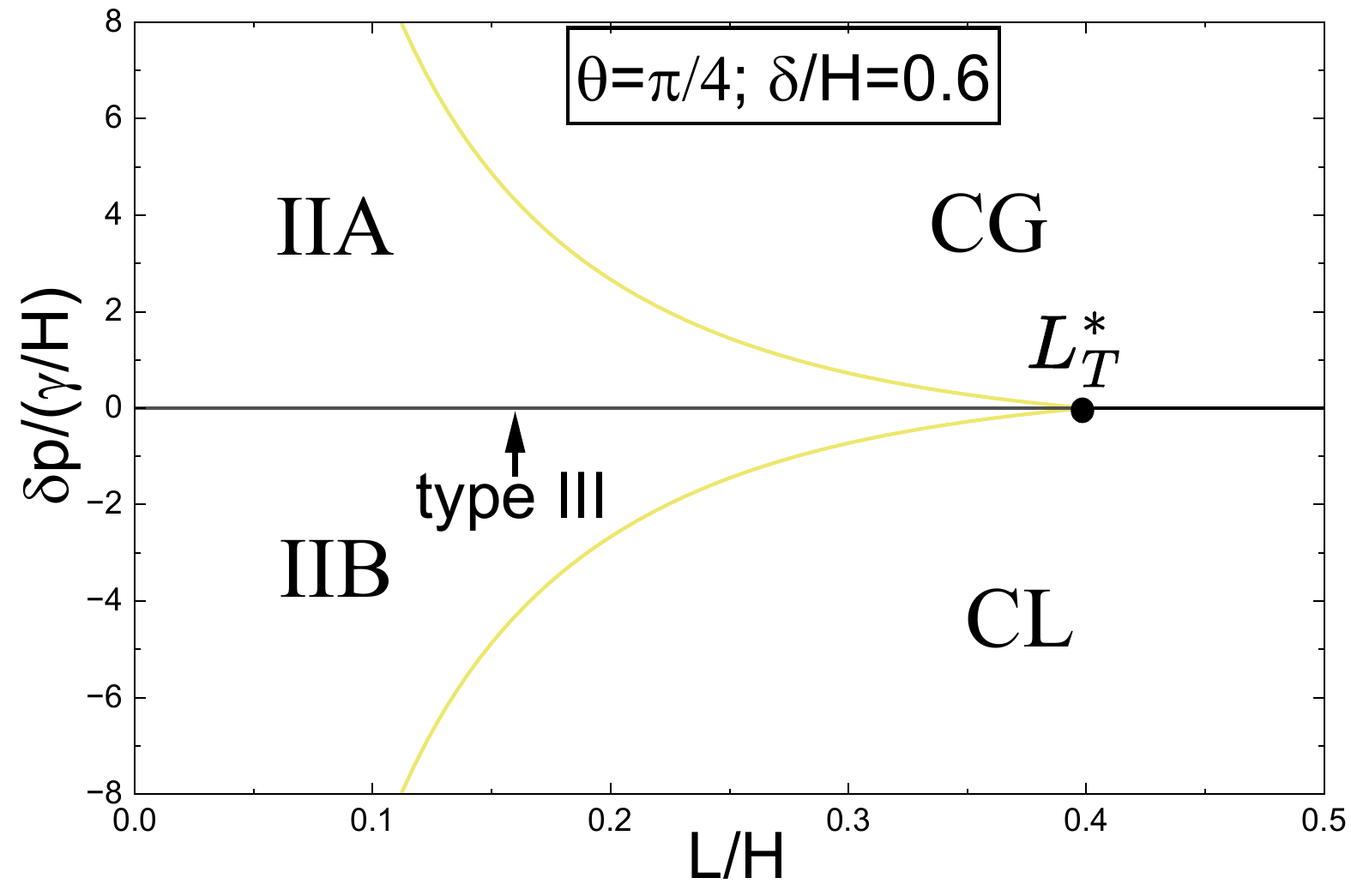}\hspace*{0.1cm} 
	\caption{Sections of the surface phase diagram showing the phase boundaries separating the stability of the capillary gas (CG), capillary liquid (CL), pinned (type I), and partially pinned (type IIA, B) bridging phases, for different slit widths and pressures: a) maximum contrast b) partial contrast with $\delta<\delta_\dagger$, c) partial contrast with $\delta>\delta_\dagger$. For partial contrast, and only at $\delta p =0$, the bridging phases are of type III and the menisci can freely slide until they meet an AB boundary.} 		
\end{figure*}

We focus on low temperatures, away from the vicinity of the bulk critical point, i.e., when $H,L$ and $\delta$ are much larger than the bulk correlation length $\xi_b$. Six fluid phases may be identified macroscopically:  a low density capillary gas (CG), a high density capillary liquid (CL) and a further four in which the adjacent stripes are connected by an array of liquid bridges separated by menisci which span the capillary. These bridging states are distinguished by how the menisci ends are pinned \cite{Abraham2007}. They may be fully pinned (bridge type I)  with each end anchored at an AB boundary and characterized by edge contact angles  $\theta_e$ and $\theta_{e}'$ \cite{Malijevsky2017a, Malijevsky2021a} -- see Fig.~1a.  In the absence of shearing this is the only possible bridging state \cite{Chmiel1994, Roecken1998, Laska2021, Abraham2024}. When $\delta \ne 0$, it is also possible that one end is a pinned at an AB boundary, while the other is unpinned and either meets stripe A with contact angle $\theta$  or stripe B with contact angle $\pi-\theta$.  These we refer to as types IIA and IIB bridges respectively -- see Fig.~1b,c. Finally, when $p=p_{\rm sat}$ it is possible that each end of the menisci is unpinned, meeting patch A at angle $\theta$ and patch B at angle $\pi-\theta$, and free to slide till they meet an AB boundary; we refer to these as type III bridges, see Fig. 1(d). The macroscopic grand potential $\Omega$ for each phase is determined by the pressure contributions from the volumes of the gas and liquid and the surface tensions arising from contact with the stripes and the menisci which are circular arcs of Laplace radius $R=\gamma/\delta p$. where $\delta p=p_{\rm sat}-p$. The phase with  lowest $\Omega$ is stable and determines the phase diagram.

Sections of the macroscopic phase diagram, showing the phase transitions as the pressure is increased (say), for different geometrical parameters, $L/H$ and $\delta/H$ are shown in Fig.~2. As with interface delocalization the value of the contact angle is crucial. We distinguish between the cases of {\it{ maximum contrast}}, for which $\theta=0$ so A is completely wet while B is completely dry, and {\it{partial contrast}} where A is partially wet and B is partially dry:\\

{\it{Maximum contrast}} (Fig.~2a). There are two characteristic slit widths,  $L_T(\delta)=\sqrt{H^2-\delta^2}$ and $L_D(\delta)$ (to be specified later), and subsequently three regimes.  For wide slits, with $L>L_T(\delta)$, condensation occurs directly from CG$\to$CL at $\delta p_{cc}=0$.  Below this value, condensation always involves two first-order transitions occurring away from bulk saturation -- a transition from CG to a bridge phase, and from a bridge phase to CL. For slits of intermediate width, $L_T(\delta)>L>L_D(\delta)$, the bridge phases are type I (fully pinned) and the transitions occur at $\pm\delta p_I$ (with the dependence on $L/H,\delta/H$ to be specified later). Intriguingly, the length-scale $L_D(\delta)$ has a maximum occurring at $\delta/H\approx 0.55$, so that the bridging phases coexisting with CG are of type I for small and large shears but of type IIA for intermediate values. For still narrower slits with $L<L_D(\delta)$ the transitions involve type IIA and type IIB bridges and occur at $\pm\delta p_{II}$. In this case, there are also two third-order meniscus depinning transitions, between type I and type II bridges, at $\pm\delta p_D$ (shown as dashed green lines).\\

\begin{figure*}
	\includegraphics[width=0.5\textwidth]{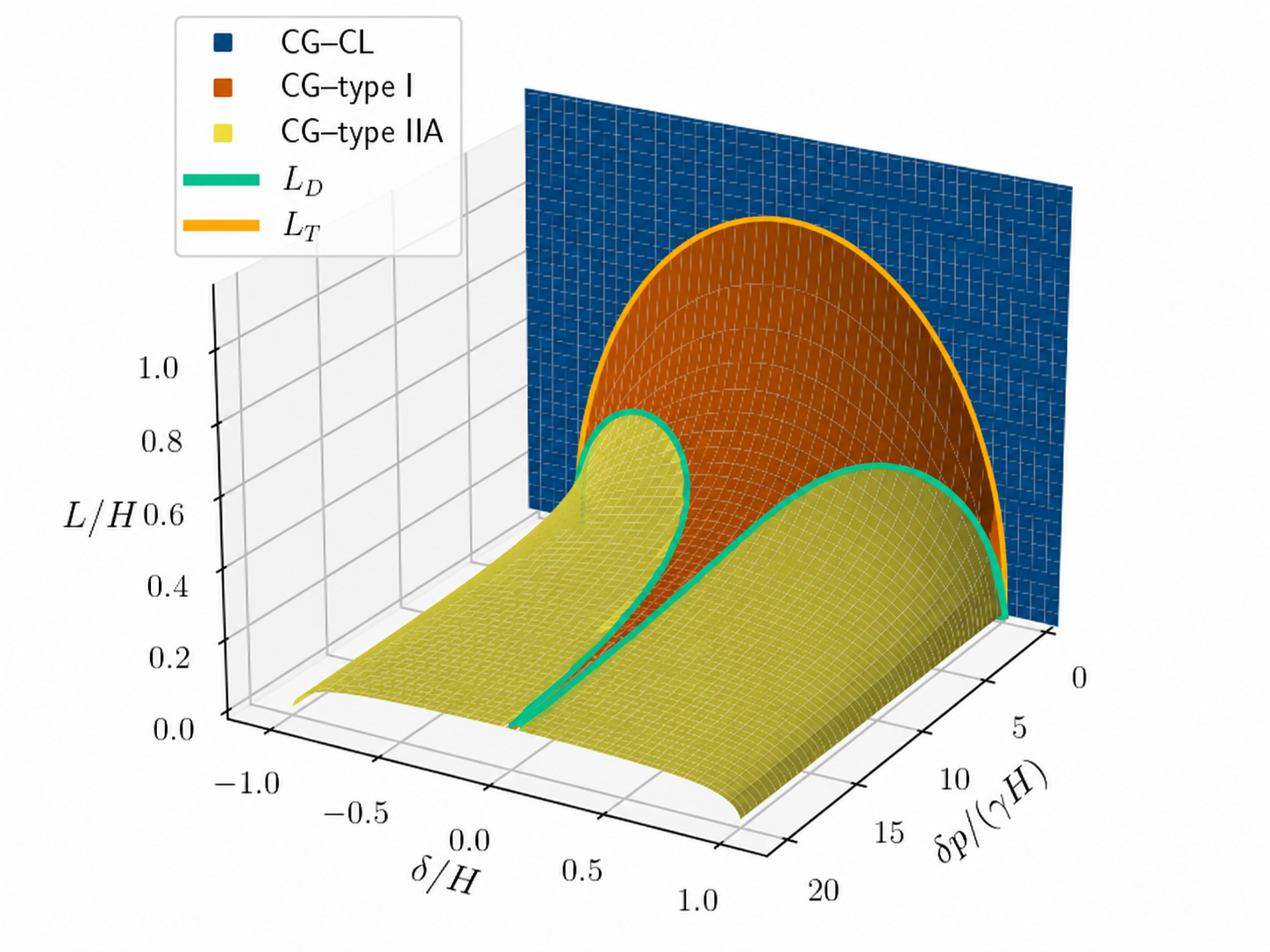}\hspace*{0.1cm} \includegraphics[width=0.5\textwidth]{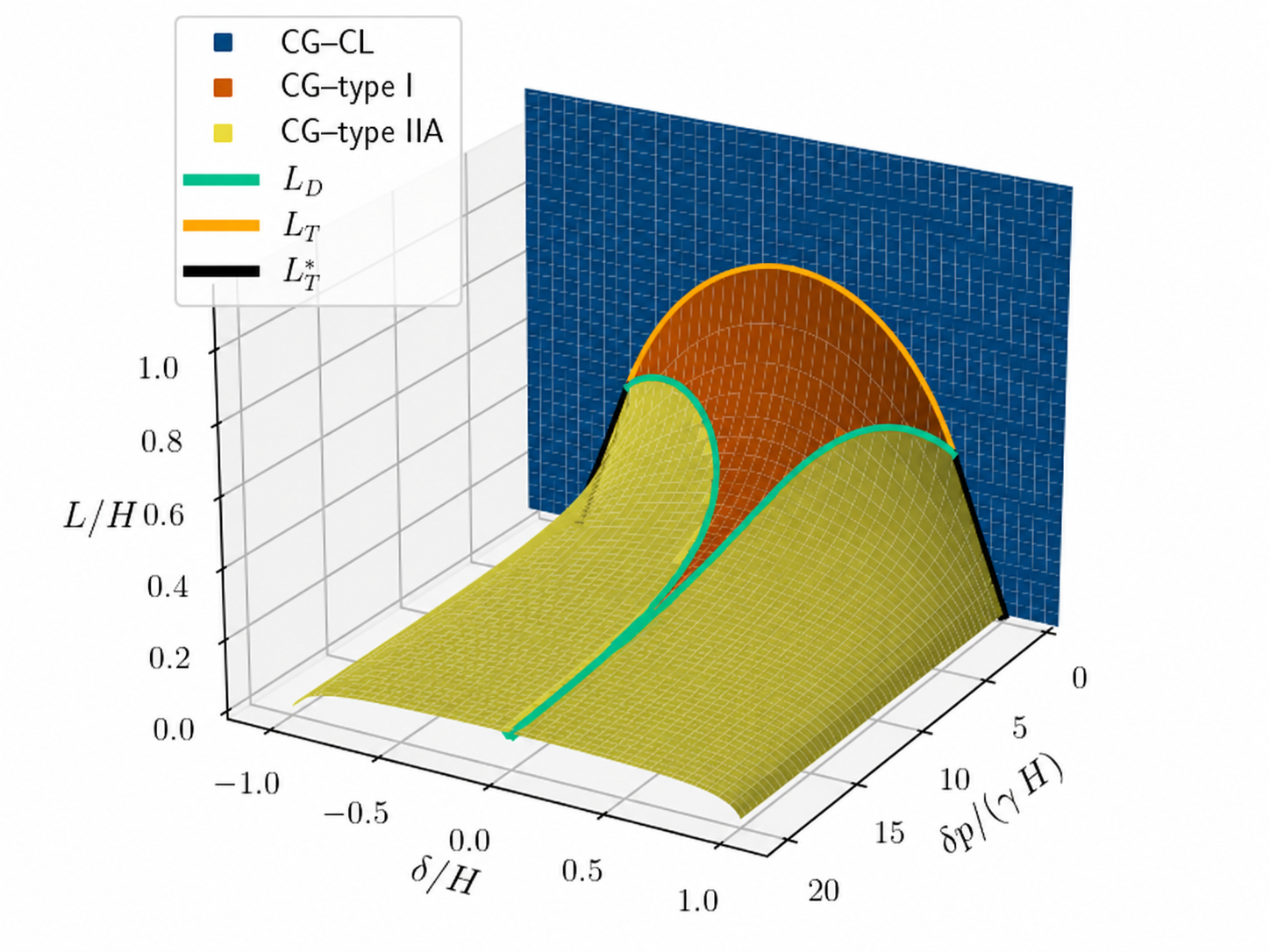}
	\caption{Three dimensional phase diagrams showing the surfaces of first-order CG$\to$CL (blue), CG$\to$ type I bridge (red) and CG $\to$ type IIA (yellow) bridge phases for maximum and partial contrast. The green line is the crossover between type I and type IIA  bridging phases. At $\delta p=0$, three-phase coexistence occurs along the orange and black lines between CG, CL and type I  and type III  bridging phases, respectively. Left: maximum contrast slit ($\theta=0$); right:  partial contrast slit ($\theta=30\degree$).}
\end{figure*}

 {\it{Partial contrast}} (Fig.~2b,c): Here the behaviour depends if the shear is smaller or larger than a threshold value $\delta_\dagger=H\cos^2\theta$. If $\delta<\delta_\dagger$, there are three characteristic widths,  $L_T(\delta)=\sqrt{H^2\cos^2\theta-\delta^2}$, $L_D(\delta)$ and $ L_{AB}(\delta)=\delta\tan\theta$, and hence four regimes. Three of these are similar to the maximum contrast case. However, for $L<L_{AB}(\delta)$ and exactly at  $\delta p=0$, the bridges are of type III, i.e., the menisci ends are unpinned and can freely slide until they meet an AB boundary.  This corresponds to a near Goldstone mode, where translations of the local meniscus position do not cost any bulk or surface free-energy.  The two lines of type I/II meniscus depinning transitions, for the regime $L_{AB}(\delta)<L< L_D(\delta)$,  are also now second-order. As the shear is increased, the three characteristic widths converge so that for  $\delta>\delta_\dagger$, there is only one characteristic width $L_T^*(\delta)=(H-\delta)\cot\theta$ and hence only two regimes: When $L>L_T^*(\delta)$, CG condenses to CL at $\delta p_{cc}=0$, while for $L<L_T^*(\delta)$ and $\delta p=0$, the bridging phase is type III. Since the menisci are laterally delocalized, the associated interfacial fluctuations and correlation length of type III bridges are much larger than those for type I and type II. For example, the present interfacial analysis predicts that, for systems with short-range forces, the correlation length along the direction of translational invariance, $\xi_y,$  is of the order $\xi_y\approx \exp[\delta\sin\theta/4\xi_b]$.

 The corresponding three-dimensional phase diagrams, showing the surfaces of first-order phase boundaries in $(\delta p, L/H,\delta/H)$ space are shown in Fig.~3a (maximum contrast) and Fig.~3b,c (partial contrast) illustrated, here for the case $\theta=\pi/6$. The phase diagram is periodic outside the interval $-1<\delta/H<1$, and is also shown only for $\delta p>0$ describing the transitions of the CG. A symmetrical structure describes the transitions for the CL, occurring for $\delta p<0$. The blue sheet represents the simple, one-step, capillary condensation from CG $\to$ CL, at $\delta p_{cc}=0$, while the red and yellow sheets are the surfaces of CG $\to$ type I and CG $\to$ type IIA bridging respectively. These meet along the green line, which is the intersection with the surface of continuous meniscus depinning transitions, which occur closer to bulk saturation. At $\delta p =0$, the line of triple points, described by $L_T(\delta)=\sqrt{H^2\cos^2\theta-\delta^2}$ where CG, CL and type I bridges coexist, is shown in orange. For partial contrast, the curve $L_T^*(\delta)=(H-\delta)\cot\theta$, (black), represents the line of triple points where CG, CL and type III (unpinned bridging phase) coexist.

Direct condensation from CG $\to$ CL, and the formation of type III bridges, both occur at bulk saturation, $\delta p=0$,  since the average (Cassie) value of the contact angle at the walls is $\pi/2$ \cite{Cassie1948}. The location of the other phase transitions surfaces of coexistence, are described by generalised Kelvin equations. The first-order transition from CG $\to$ type I, fully pinned, bridge phase occurs when 

\begin{equation}
	\delta p_I=\frac{2\gamma \cos\phi \cos\alpha}{L}\,,
\end{equation}
where $\tan\alpha=\delta/L$ is the shear angle, and $\phi=\theta_e-\alpha$, is obtained implicitly from
\begin{equation}
	\cos\theta=\cos\phi\cos\alpha+\frac{L\sec\alpha}{2H}\left[\sin\phi+\sec\phi\left(\frac{\pi}{2}-\phi\right)\right]\,.
\end{equation}
Here $\theta_e$ and $\theta_e’=\theta_e-2\alpha$ are the values of the edge contact angle at the bottom and top walls respectively. The phase boundary for the CG $\to$ type II A transition, is given by
\begin{equation}
	\delta p=\frac{\gamma(\cos\theta+\cos\theta_e)}{L}\,,
\end{equation}
where now the edge contact $\theta_e$ takes a different value, determined implicitly from
\begin{equation}
	\cos\theta=\cos\theta_e+\frac{L}{d}\frac{\left[\sin(\theta+\theta_e)+\pi-\theta-\theta_e\right]}{\cos\theta+\cos\theta_e}\,,
\end{equation}
where $d=H-\delta+L\tan(\theta_e-\theta)/2$ is the length of the A patch covered by liquid. Lastly, the surface of meniscus depinning transitions, whose cross-sections give the green dashed lines in Fig 2a,b occurs at
\begin{equation}
	\delta p_D=\frac{2\gamma(L\cos\theta-\delta\sin\theta)}{L^2+\delta^2}
	\end{equation}
	Setting $\delta p_D=0$ gives the characteristic width $L_{AB}(\delta)$. Similarly the condition $\delta p_I=\delta p_{II}=\delta p_D$ determines the solid green crossover line where the three phase boundaries meet. This also gives the remaining characteristic width $L_D(\delta)$, determined implicitly by a transcendental equation,
which is equivalent to the condition $d=H$. We note that the crossover line has a qualitative change at $\theta=\pi/4$, and develops a maximum for smaller values of the contact angle.

\begin{figure*}
	\includegraphics[width=0.5\textwidth]{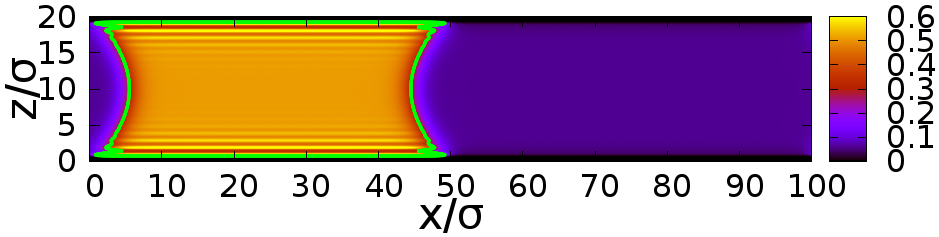}\hspace*{0.1cm} \includegraphics[width=0.5\textwidth]{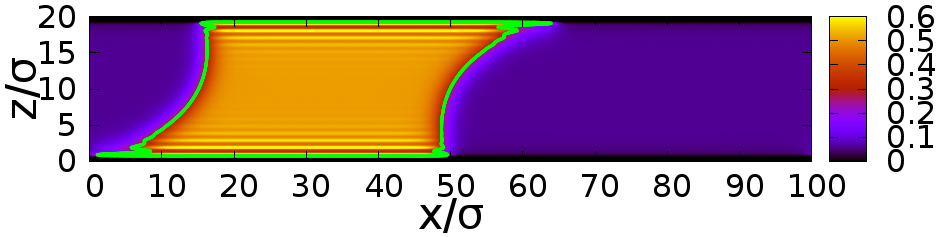}
	\includegraphics[width=0.5\textwidth]{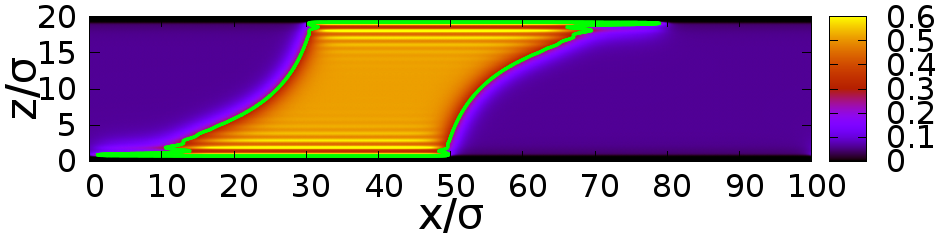}\hspace*{0.1cm} \includegraphics[width=0.5\textwidth]{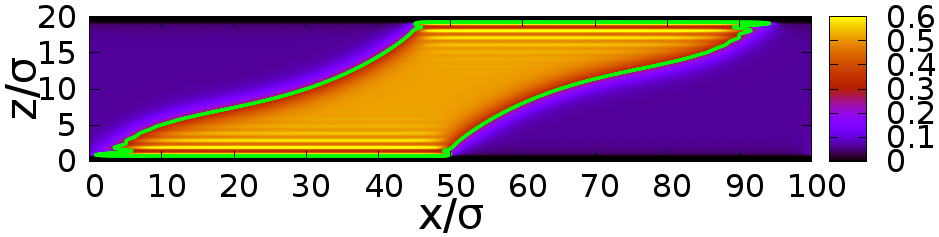}
	\caption{	
		DFT equilibrium density profiles for a maximum contrast slit with $H=50\,\sigma$,  $L=20\,\sigma$ and increasing shear displacements. Each of the these bridging states coexist with a CG phase. For $\delta=0$, $15$, and $45\,\sigma$ the bridging is of type I involving pinned menisci. For the intermediate shear with $\delta=30\,\sigma$, the bridging phase is of type II, and only one end is pinned. For the largest shear with $\delta=45\,\sigma$ which corresponds to the triple point where CG, CL and type I bridge coexist, the menisci remain strongly curved, even though the pressure is very close to bulk saturation demonstrating deviations from the macroscopic expectation due to the influence of intermolecular forces.}
\end{figure*}

\begin{figure}
	\includegraphics[width=0.5\textwidth]{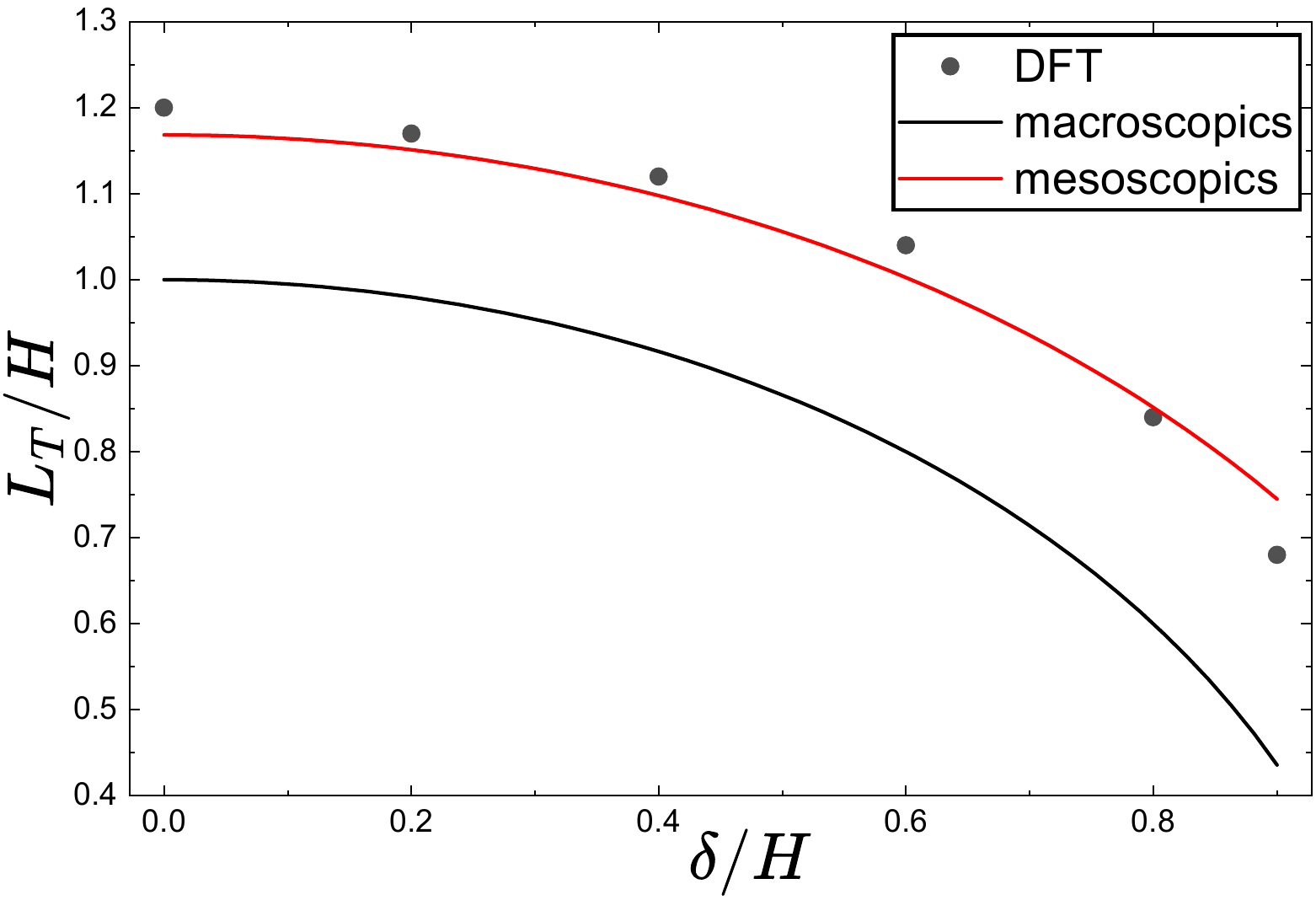} 
	\caption{
		Triple-point line $L_T(\delta)$ for the maximum contrast case ($\theta=0$) showing the comparison between the macroscopic, mesoscopic and microscopic DFT results.  Mesoscopic results account for long-ranged intermolecular forces which affect the surface free-energy of wetting drops and drying bubbles on the A and B patches. }\label{fig_lmax}
\end{figure}

These predictions for the phase equilibria and associated phase boundaries remain accurate below the capillary critical point, i.e. provided $\xi_b/L\ll1$ \cite{Privman1983}. In this case the condensation from CG to CL, or from CG to either of the bridging phases, is sharply first-order, while the first, second and third-order transitions between the bridging states are rounded only over a very narrow scale set by $\xi_b/\delta$. For very narrow capillaries, effects associated with molecular layering will also become apparent. For maximum contrast, where complete wetting and drying films amplify dispersion-force effects, mesoscopic corrections arise in two ways. Firstly, due to the  presence of thick wetting and drying layers on the patches, which in turn reveals the influence of long-ranged, dispersion-like, intermolecular forces. These are most important for the phase coexistence occurring near bulk saturation, i.e., along the line of triple points where CG, CL, and type I bridges coexist, where the wetting layers are thickest. For example, the CG phase will have a droplet of liquid on each A patch, with ends pinned at each AB boundary, and similarly the CL phase will have bubbles of gas adsorbed on each B patch. These drops and bubbles introduce logarithmic, size-dependent corrections to the surface free-energy, analogous to a Casimir contribution to the line tension \cite{Abraham2010, Malijevsky2017}.  Allowing for this, the macroscopic prediction for the the triple line, $L_T(\delta)=\sqrt{H^2-\delta^2}$, is modified to
\begin{equation}	
	L_T(\delta)\approx\sqrt{(H+C\ln (H/\sigma))^2-\delta^2}\,,	\label{Lmax_meso}
\end{equation}
where $\sigma$ is the molecular diameter and is another microscopic length determined as $C=\sqrt{2c_A/\gamma}+\sqrt{2c_B/\gamma}$, where $c_A$ and $c_B$ are the Hamaker constants for the binding potentials at the A and B patches. The location of the triple line of three-phase coexistence should therefore lie slightly above the macroscopic prediction. Secondly, mesoscopic effects are important for the type I bridging phase where each meniscus spans the capillary. At a purely macroscopic level, the pinned menisci are simply straight lines, since the triple line occurs at bulk saturation $p=p_{\rm sat}$. However, this does not allow for the repulsion of the interface from both walls reflecting the onset of interfacial delocalization, which becomes increasingly important as the aspect ratio $\delta/H$ approaches unity.  Instead of being a straight line, the interfacial height $\ell(x)$ describing the meniscus shape,  is then described by a scaling expression
\begin{equation}
	\ell(x)\approx \frac{L}{2}\left[1+\frac{\sinh((x-H-\delta)/\xi_y)}{\sinh((H+\delta)/2\xi_y)}\right]\,,\label{xi_par}
\end{equation}
where $x$ is the coordinate along the lower wall, with the origin placed at an AB boundary. The correlation length $\xi_y$ characterises fluctuations along the direction of translational invariance $y$ and scales as  $\xi_y\sim \exp[L/4\xi_b]$ for systems with short-range forces and  $\xi_y\sim L^2$ for systems with long-range forces -- these results are the same as the predictions for interfacial delocalization in capillary slits with walls of opposing wetting preferences \cite{Parry1992}.

\begin{figure}
	\includegraphics[width=0.5\textwidth]{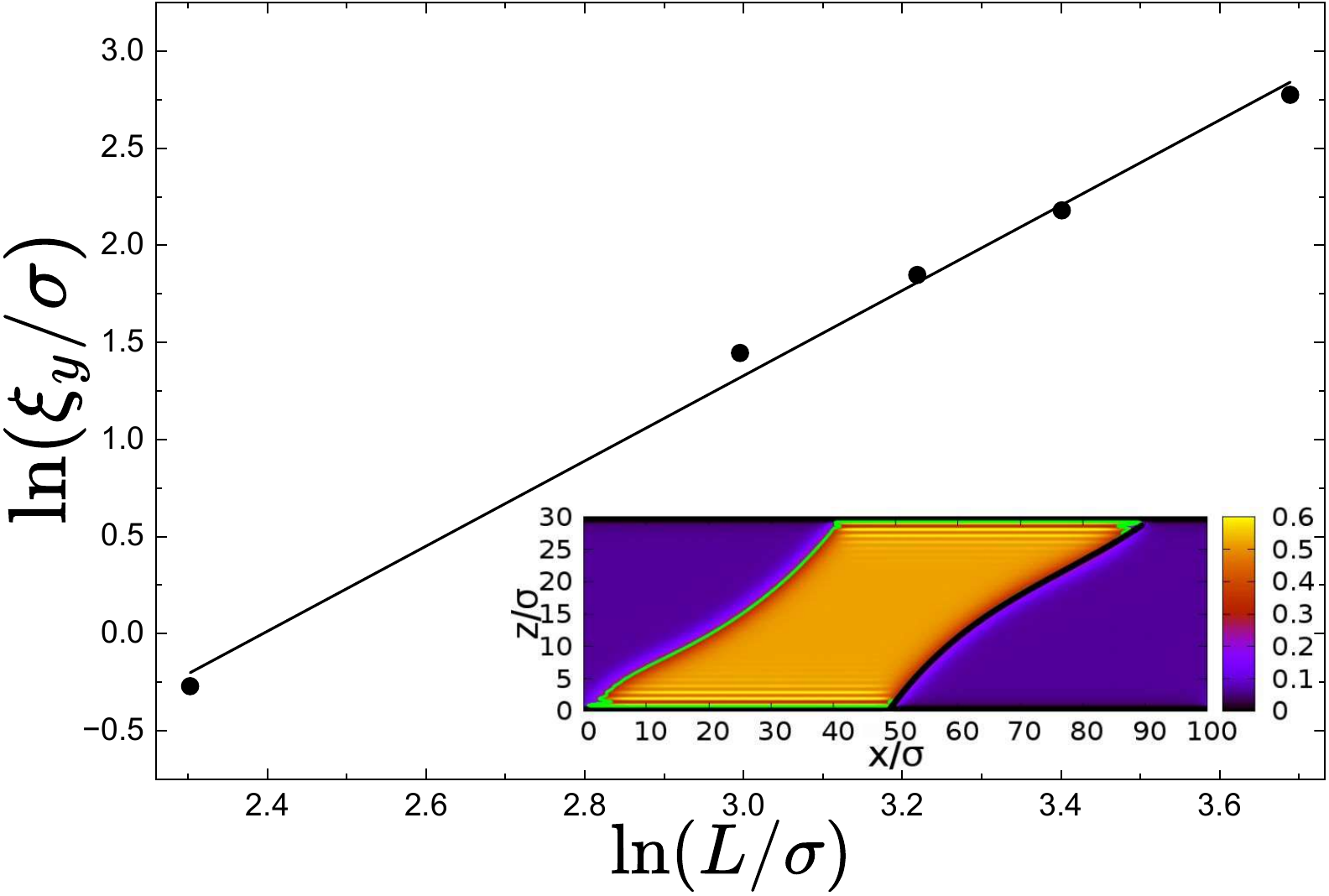} 
	\caption{
		The correlation length $\xi_y$ obtained from the DFT results for the maximum contrast slit with  $\delta=40\,\sigma$, and $H=50\,\sigma$ and different slit widths.  The log-log plot gives a slope $2.19$, close to the expected scaling $\xi_y\sim L^2$. Inset: DFT density profile for $L=30\,\sigma$ with the interface fitted by Eq.~(\ref{xi_par}), from which $\xi_y$ is extracted.} \label{fig_xi}
\end{figure}

To test these predictions for the maximum contrast slit, we used a microscopic density functional theory (DFT), in order to obtain the equilibrium density profiles and grand potentials of stable and metastable phases \cite{Evans1979}. These are obtained by minimizing a grand potential, which is the functional of density profile $\rho(x,y)$,
\begin{equation}
	\Omega[\rho]=F[\rho]-\int d\rr\rhor\left[V(\rr)-\mu\right]\,,
	\end{equation}	
where $V(\rr)$ is the external field modelling the slit patterning, which includes long-ranged dispersion forces. Here, $F[\rho]$ is the Helmholtz free energy functional, for which we use  Rosenfeld's fundamental measure theory \cite{Rosenfeld1989} to model volume exclusion, together with a mean-field treatment of the attractive fluid-fluid interactions (see Supplementary Material \cite{SM}). Fig.~4 shows representative equilibrium density profiles for type I and type IIA bridging phases, which coexist with CG. Notice that the bridging states change from type I to type IIA and back to type I as the shear increases --  consistent with the predicted non-monotonic behaviour for the characteristic length-scale $L_D(\delta)$. The DFT results  for the triple line $L_T(\delta)$, where CG, CL and type I bridges coexist, are shown in Fig.~5. The comparison demonstrates the accuracy of the mesoscopic prediction \eqref{Lmax_meso} and shows explicitly how the logarithmic correction accounting for the influence of the intermolecular forces shifts the triple line above the purely macroscopic result.  Finally, in Fig.~6 we show DFT results demonstrating the growth of $\xi_y$ with $L$ and (inset) the accuracy of the scaling expression  \eqref{xi_par} for the density profile which reveals deviations from the macroscopic prediction that the menisci should be planar.

In this Letter we have shown that fluids confined between chemically patterned walls exhibit a remarkably rich phase behaviour  due to the presence of bridging phases which have different pinning properties. The resulting phase diagrams display multiple first-order and continuous transitions, as well as multiphase coexistence, captured by generalised Kelvin equations with mesoscopic corrections and confirmed by fully microscopic DFT. We expect that similar phenomena will also occur in geometrically patterned confinements. In more complex geometries, particularly for incommensurate periodicities or imperfect chemical contrast, the number of competing bridging states is expected to increase markedly, with a shallow accompanying free-energy landscape providing a crossover between the phase equilibria of ordered patterned slits and the interfacial behaviour of disordered confinements. We hope that these predictions will stimulate experimental studies of adsorption, bridging, and interfacial pinning in patterned confinements.

\end{document}